\documentclass[preprint2]{aastex}

\newcommand{\shape}{{\it Shape\ }}
\newcommand{\Mdot}{\dot{M}~}
\newcommand{\kms}{\mbox{ km s$^{-1}$}~}
\newcommand{\Moy}{\mbox{M$_{\odot}$ yr$^{-1}$}~}

\shorttitle{Hydrodynamical Velocity Fields in Planetary Nebulae}
\shortauthors{Steffen and Garc\'{\i}a-Segura}

\begin{document}

\title{Hydrodynamical Velocity Fields in Planetary Nebulae}

\author{ Wolfgang Steffen, }
    \affil{Instituto de Astronom\'{\i}a,
    Universidad Nacional Aut\'onoma de M\'exico,
    Ensenada, B.C., M\'exico}
    \email{wsteffen@astrosen.unam.mx}

\author{
  Guillermo Garc\'{\i}a-Segura,}
    \affil{Instituto de Astronom\'{\i}a,
    Universidad Nacional Aut\'onoma de M\'exico,
    Ensenada, B.C., M\'exico}
    \email{ggs@astrosen.unam.mx}

\and

\author{
  N. Koning}
  \affil{Space Astronomy Laboratory, University of Calgary, Canada.}
  \email{nkoning@iras.ucalgary.ca}

\begin{abstract}
Based on axi-symmetric hydrodynamical simulations and 3D reconstructions with {\it Shape},
we investigate the kinematic
signatures of deviations from homologous (``Hubble-type'') outflows in some typical
shapes of planetary nebulae. We find that, in most situations considered
in our simulations,
the deviations from a Hubble-type flow are significant and
observable. The deviations are systematic and a simple
parameterization of them considerably improves morpho-kinematical models of the
simulations.
We describe such extensions to a homologous expansion law that capture the
global velocity structure of hydrodynamical axi-symmetric nebulae during their
wind-blown phase. It is the size of the poloidal velocity component that strongly
influences the shape of the position velocity diagrams that are obtained,
not so much the variation of the radial component.
The deviations increase with the degree of collimation of the nebula and
they are stronger at intermediate latitudes.
We describe potential deformations
which these deviations might produce in 3D reconstructions
that assume ''Hubble-type'' outflows. The general conclusion
is that detailed morpho-kinematic observations and modeling of planetary nebulae
can reveal whether a nebula is still in a hydrodynamically active stage
(windy phase) or whether it has reached ballistic expansion.
\end{abstract}


\section{Introduction}

\label{sec:intro}

Most methods used for the reconstruction of three dimensional
structures of planetary nebulae are based on imaging and kinematic
data. Usually, they assume homologous expansion, i.e., purely radial velocity
with magnitude proportional to the distance from the
central star (e.g. Sabbadin et al. 2006; Santander-Garc\'{\i}a et al. 2004;
Hajian et al. 2007; Steffen \& L\'opez 2006;
Morisset et al. 2008; Dobrin{\v c}i{\'c} et al. 2008).
The general justification for adopting this so called
"Hubble-type expansion" is that observed expansion speeds directly
correlate with distance from the center of the nebula
(Wilson 1950, O'Connor et al. 2000, Meaburn et al. 2008). Furthermore, the structures
seen in position-velocity (P-V) diagrams are often very similar to the  direct
image
(e.g. Mz-3: Santander-Garc\'{\i}a 2004; or NGC6302: Meaburn et al. 2005).
A technical reason for adopting this type of expansion law
is that, it is the simplest direct mapping between
the velocity of a gas
parcel and its position along the line of sight.
Still, this method requires an additional constraint
to determine the constant of
proportionality that links the Doppler-velocity to the position along
the line of sight. This constraint may be an estimate of the extent
of the object along the line of sight as deduced from partial symmetries
or other arguments.

The homologous expansion
is a reasonable approximation if the motion of the gas has become
ballistic only a short time after the leaving the center
of the nebula as compared to the current age of the nebula (for a discussion
of this issue see e.g. Steffen \& L\'opez 2004). Observational and theoretical evidence is mounting that
the hydrodynamical events that determine the structure formation in many planetary nebulae are rather short-lived (e.g. Alcolea, Neri \& Bujarrabal 2007; Meaburn et al. 2005; Dennis et al. 2008; Akashi \& Soker 2008). In order to decide whether the extended interacting wind scenario can be held up as it has been proposed almost two decades ago, it becomes important to look for kinematic evidence.
During the action of the interacting winds, important deviations from a Hubble-type expansion law are to be expected in the velocity magnitude and direction.

We study the kinematics of hydrodynamical simulations as a base to look
for such deviations in observed nebulae. This study also aims to improve
three-dimensional modeling that is being
carried out with photo-ionization codes (e.g. Morisset 2006; Ercolano, Barlow \& Storey 2005) and morpho-kinematic reconstruction methods (e.g. Sabbadin et al. 2006; Steffen \& L\'opez 2006).

The morpho-kinematical modeling code \shape has been designed to allow the 3D modeling
of very complex emissivity structures and velocity fields (Steffen \& L\'opez 2006).
{\it Shape}, and most current photo-ionization codes, do not actually calculate the structure
and velocities from physical initial conditions. It is therefore important to have a physical guidance when deviations from a homologous expansion are to be included in morpho-kinematical modeling.
To provide improved velocity fields we study the kinematics of some typical morphologies of planetary nebulae as created by two different mechanisms, purely hydrodynamics and magnetic confinement.

As discussed by Li, Harrington \& Borkowski (2002) detailed kinematic models are required to improve the accuracy of distance measurements to planetary nebulae from kinematic data. As these authors stress, deviations from homologous expansion should be taken into account to improve accuracy. Our paper aims at validating the \shape code for later application to the type of data presented by Li, Harrington \& Borkowski (2002) for BD+30$^{o}$3639.

First, we apply purely hydrodynamical numerical simulations to create elliptical and bipolar nebulae from an equatorial density enhancement in the environment. Highly elongated bipolar nebulae are simulated with magneto-hydrodynamic calculations.
The kinematics of the simulated data is analyzed and then reproduced as a morpho-kinematical model using {\it Shape}.
We assess the distortions that might occur in 3D reconstructions of such objects.
In addition to providing template velocity fields, the direct comparison of the simulations
with the  reconstructed models using \shape allows us to validate the modeling
capabilities of \shape itself.

In Section 2 we describe the hydrodynamical model setup.
Section 3 contains a description of the morpho-kinematical modeling technique and in Section 4 we discuss and summarize our results.

\section{Numerical simulations}
\label{sec:models}

The four reference simulations have been performed using the
magneto-hydrodynamic code ZEUS-3D (version 3.4), developed by M. L. Norman
and the Laboratory for Computational Astrophysics.
This is a finite-difference, fully explicit,
Eulerian code descended from the code described in Stone \& Norman (1992).
A method of characteristics is used to compute magnetic fields as described
in Clarke (1996).
The models include the Raymond \& Smith (1977) cooling curve above $10^4$ K.
For temperatures below $10^4$ K, {\bf the radiation cooling term is set to
zero}.
Photoionization is not included for simplicity.
The initial and minimum temperature allowed in all models is set to $10^2$ K.

The computations are done in spherical coordinates (r, $\theta$),
with outflowing outer boundary conditions  and reflective conditions at the
pole and equator.
The models have grid resolutions of $200 \times 180 $ in r and $\theta$,
with a radial extent of 0.1 pc and 90$^\circ$, respectively.
The inner radial boundary is located at $2.5 \times 10^{-3}$ pc, i.e.,
we skip the inner $2.5 \%$ of the computational mesh.

The winds are set based on the rotating wind solutions given by
Bjorkman \& Cassinelli  (1993), Ignace et al. (1996)
and the approach to those equation given by Garc\'{\i}a-Segura et al. (1999).
These equations permit the introduction of the winds self-consistently.
The  magnetic field of an outflowing wind from a rotating star
can be described by two components, $B_{\phi}$ and $B_{\rm r}$
(Chevalier  \& Luo 1994; R\'o\.zyczka \& Franco 1996; Garc\'{\i}a-Segura
et al. 1999).
The radial field component  can be neglected since $B_{\rm r}
\sim r^{-2}$ while the toroidal component $B_{\phi} \sim r^{-1}$, and the
field configuration obeys $ \nabla \cdot B = 0 $ .

Two useful dimensionless parameters
containing the basic information of the models are used.
The first one, $\Omega$, is the velocity ratio of the stellar rotation to the
critical rotation, and the second parameter, $\sigma$, is the ratio of
the magnetic field energy density to the  kinetic energy density in the
wind.
As a initial condition, the AGB wind is set to fill up the whole
computational mesh, while the fast wind is set only at the inner
radial boundary, using equation (2) and (3) in Garc\'{\i}a-Segura et al.(1999)
for both winds with their respective parameters. Note that equation (2)
only sets the radial component of the wind velocity, since $v_{\phi}$
is negligible (at the inner boundary distance) due to angular momentum
conservation.

\begin{figure}[!t]\centering
  \vspace{0pt}
  \includegraphics[angle=00,width=0.9\columnwidth]{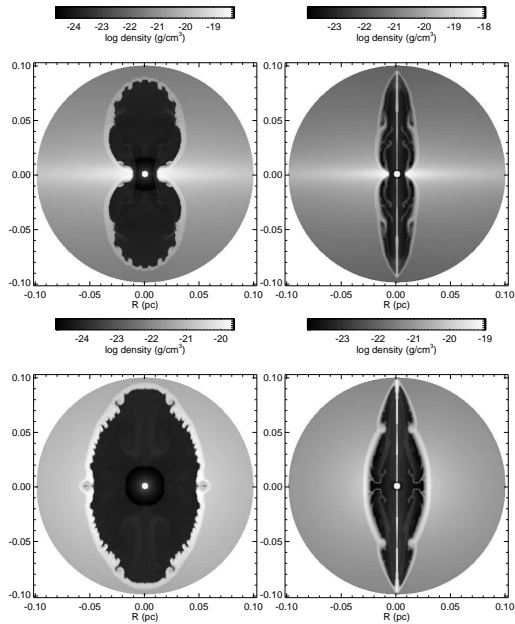}
  \caption{Density distributions of model B at 2500 yr (top-left),
        model E  at 3500 yr (bottom-left),
        model Bm at 1500 yr (top-right), and
        model Em at 2000 yr (bottom-right).}
  \label{fig:simulations}
  \vspace{0pt}

  \end{figure}

The four runs (Figure \ref{fig:simulations})
include, a bipolar (B) and an elliptical (E) case without
magnetic fields, and a bipolar (Bm) and an elliptical (Em) case with magnetic
fields on.  The AGB slow winds have velocities of 10 \kms and
$\Mdot = 10^{-5} $ \Moy in all cases, while the fast winds have
velocities of 1000 \kms and $\Mdot = 10^{-8} $ \Moy.
Bipolar runs (B, Bm) have $\Omega=0.98$ in the AGB phases, while
elliptical ones $\Omega=0.6$.
Magnetic models have  $\sigma = 0.01$  only in  the fast wind.
AGB winds are set unmagnetized (see Garc\'{\i}a-Segura et al. 1999 for
further details).

\section{Morpho-kinematical modeling with SHAPE}
\label{sec:shape}

In this study, the hydrodynamical simulations serve two purposes.
First, we use them as observable objects, of which we know
their structure, dynamics and kinematics. For a given time in the
evolution, we analyze and visualize the projected images and
position velocity (P-V) diagrams as if they where true observations.
These images and P-V diagrams are generated with \shape after the
data from the simulations have been imported in a suitable format (see below).

The second purpose of using hydrodynamical simulations is to validate the
functionality of \shape itself. For this purpose we reproduce the velocity
field of the simulation and its structure in detail using the structure
and velocity modeling tools of \shape. We then check
whether the images and P-V diagrams are consistent with those of the
simulation.

We consider only the dense expanding
shell that has been swept up by the fast stellar wind. It is this
shell, that is expected to be visible as the bright main shell in
a planetary nebula. We assume that the nebula is fully ionized and that
the emission measure, proportional to the square of the density,
provides the relative brightness.

In order to extract the region of the expanding shell we have
selected cells from the hydrodynamic simulation according to
their radial velocity and density contrast with neighboring cells.
The selection is based on the fact that the shell is faster
than the external medium (AGB-wind), but slower and denser than
the active fast stellar wind.

In {\it Shape}, each cell corresponds to a particle with properties
of position, velocity and brightness.
In order to convert the axi-symmetric two-dimensional hydro-data
into a 3-D spatial particle structure, 18 particle copies for each
extracted cell have been distributed randomly along the corresponding
circle around the cylindrical symmetry axis. The velocity components
of the particles have been adjusted according to the rotated position.
To visualize, analyze and further process the hydrodynamic simulations
we import the simulated particle data into \shape
(Version $2.7$).

Since Version 1.0, \shape has an integrated tool for the creation of
3-D structures and their velocity field. The simulated planetary nebula
will be used to validate our modeling software, its method and accuracy.
The hydrodynamic simulations
that we are processing have been performed in spherical coordinates.
Because of the cylindrical symmetry, only the radial and the poloidal
velocity components are non-cero. \shape has two different modeling
modes for the velocity components. One is providing an analytic formula
and the other is an editable curve, in which any number of
linear segments can be used to describe the magnitude of each velocity
component ($v_r, v_{\theta}$) as a function of the corresponding
space coordinate ($r$, $\theta$).

\begin{figure}[!t]\centering
  \vspace{0pt}
  \includegraphics[angle=00,width=0.9\columnwidth]{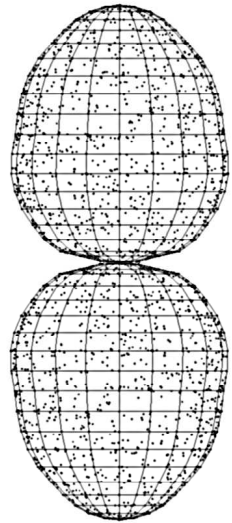}
  \caption{The surface model mesh that has been adjusted to the extracted shell from run B.
  The dots that are plotted represent only 5 \% of the total particles that have been
  stochastically applied to sample the object.}
  \label{fig:model_mesh}
  \vspace{0pt}

  \end{figure}

The main shell of
the nebula is modeled as a closed surface with cylindrical symmetry
using the surface modeling tools in \shape. As an example,
Figure \ref{fig:model_mesh} we show a visualization of the surface adjusted
to run B of the simulations. The surface density of particles is constant within the
deviations from the random particle statistics. About $6\times 10^{4}$ particles have
been used for each object.
In order to model the total emission, the relative particle emissivity
is given as a function of position with a similar tool described
for the velocity modeling (see above).

For the rendering of the images,
the emission is directly integrated along the line of sight.
In this work we assume optically thin emission. The resulting raw
image is then convolved with a gaussian point-spread-function that models the
seeing conditions of an optical observation or the beam of a radiotelescope.

Similarly, a P-V diagram is generated by distributing the emission
contained in a simulated spectrograph slit in a position-velocity image
(velocity is horizontal and position along the slit is vertical).
The raw image is then convolved with an ellipsoidal gaussian that
has width and height corresponding to the velocity and spatial resolution
of the simulated instrument.

\begin{figure}[!t]\centering
  \vspace{0pt}
  \includegraphics[angle=-90,width=0.9\columnwidth]{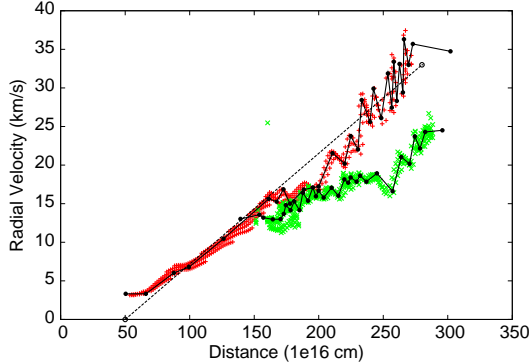}
  \caption{The radial velocity component of the ellipsoidal and bipolar lobed hydrodynamical simulations (E \& B)is plotted against the distance from the center. The lines represent the piecewise linear representation of the velocity field in {\it Shape}. The dotted line is a simplified approximation with a single linear curve. The simulated data marked by (+) show those of model B and those marked as (x) are from model E.}
  \label{fig:run1_2_r_vr}
  \vspace{0pt}

\end{figure}

\begin{figure}[!t]\centering
  \vspace{0pt}

  \includegraphics[angle=-90,width=0.9\columnwidth]{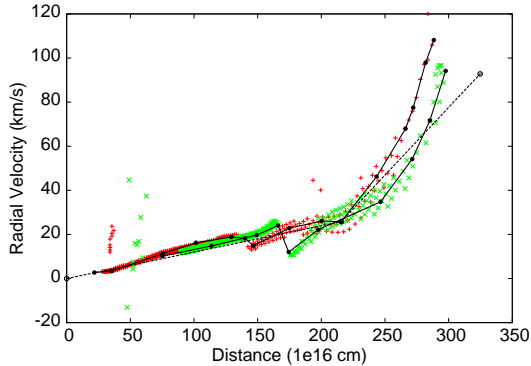}
  \caption{The radial velocity component of the MHD-simulations is plotted against the distance from the center. For a description of the lines, see Figure (\ref{fig:run1_2_r_vr}). Here, two linear curve segments have been adopted for the simplified velocity field. The simulated data marked by (+) show those of model Bm and those marked as (x) are from model Em.}
  \label{fig:run3_4_r_vr}
  \vspace{0pt}

  \end{figure}

\begin{figure}[!t]\centering

  \includegraphics[angle=-90,width=0.9\columnwidth]{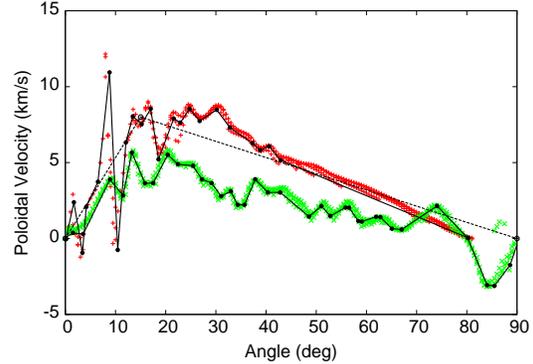}
  \caption{The poloidal velocity component of the ellipsoidal and bipolar lobed hydrodynamical simulations is plotted against the angle in degrees from the symmetry axis (equator = $90^o$). For a description of the lines and symbols, see Figure (\ref{fig:run1_2_r_vr}). Here, two linear curve segments have been adopted for the simplified velocity field.}
  \label{fig:run1_2_dt}
  \vspace{0pt}

  \end{figure}

\begin{figure}[!t]\centering
  \vspace{0pt}

  \includegraphics[angle=-90,width=0.9\columnwidth]{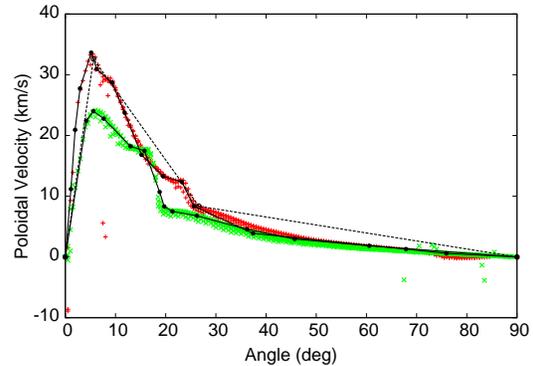}
   \caption{The poloidal velocity component of the MHD collimated simulations is plotted against the angle in degrees from the symmetry axis (equator = $90^o$). For a description of the lines and symbols, see Figure (\ref{fig:run3_4_r_vr}). Here, three linear curve segments have been adopted for the simplified velocity field.}
  \label{fig:run3_4_dt}
  \vspace{0pt}

\end{figure}

\section{Results and discussion}

In order to search for systematic tell-tale signatures which
might reveal information about the physics of their initial formation
process, we have investigated the kinematics of typical
expanding planetary nebulae based on 4 hydrodynamic simulations (Figure \ref{fig:simulations}).

Figures \ref{fig:run1_2_r_vr} and \ref{fig:run3_4_r_vr} show the distribution of the radial velocities ($v_r$) of the four simulations
as a function of distance from the center. Figures \ref{fig:run1_2_dt} and \ref{fig:run3_4_dt} display the poloidal velocity ($v_\theta$) component as a function of the angle from the polar axis of the simulations. For software validation the graphs have been approximated in \shape using piece-wise linear segments as shown by the continuous lines with dots superimposed on the data from the simulations. The resulting P-V diagrams follow those of the simulation to within the scatter of the points from the simulation, validating the our \shape code.

\begin{figure}[!t]\centering

  \includegraphics[angle=00,width=0.95\columnwidth]{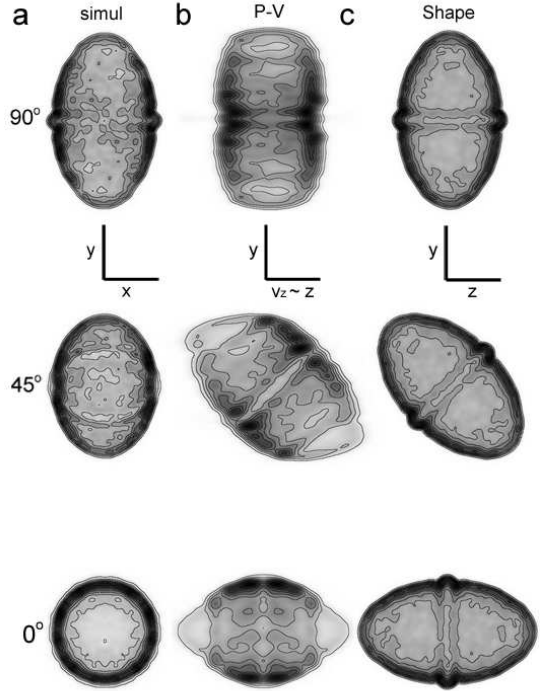}

  \caption{Column ''a''shows the rendered images (emission measure) of the hydrodynamical simulation E at different inclination angles ($0^o, 45^o, 90^o$) as seen from the observer's point of view. Column ''b'' contains the corresponding slit-less spectra of the simulation. It represents the view of the object as seen perpendicular to the line of sight in a reconstruction that is based on homologous expansion. This image reveals the deformations expected from such reconstructions. Despite of the cylindrical symmetry of the object, the shape of the ''reconstruction'' is quite different to the direct image of the simulation in column ''a''. Column ''c'' shows the rendered image of our model as reconstructed with \shape (column ''c'') with corrections to the homologous expansion velocity field as seen from the same direction as column ''b''.  }
  \label{fig:renders1}
  \vspace{0pt}

\end{figure}

\begin{figure}[!t]\centering

  \includegraphics[angle=00,width=0.95\columnwidth]{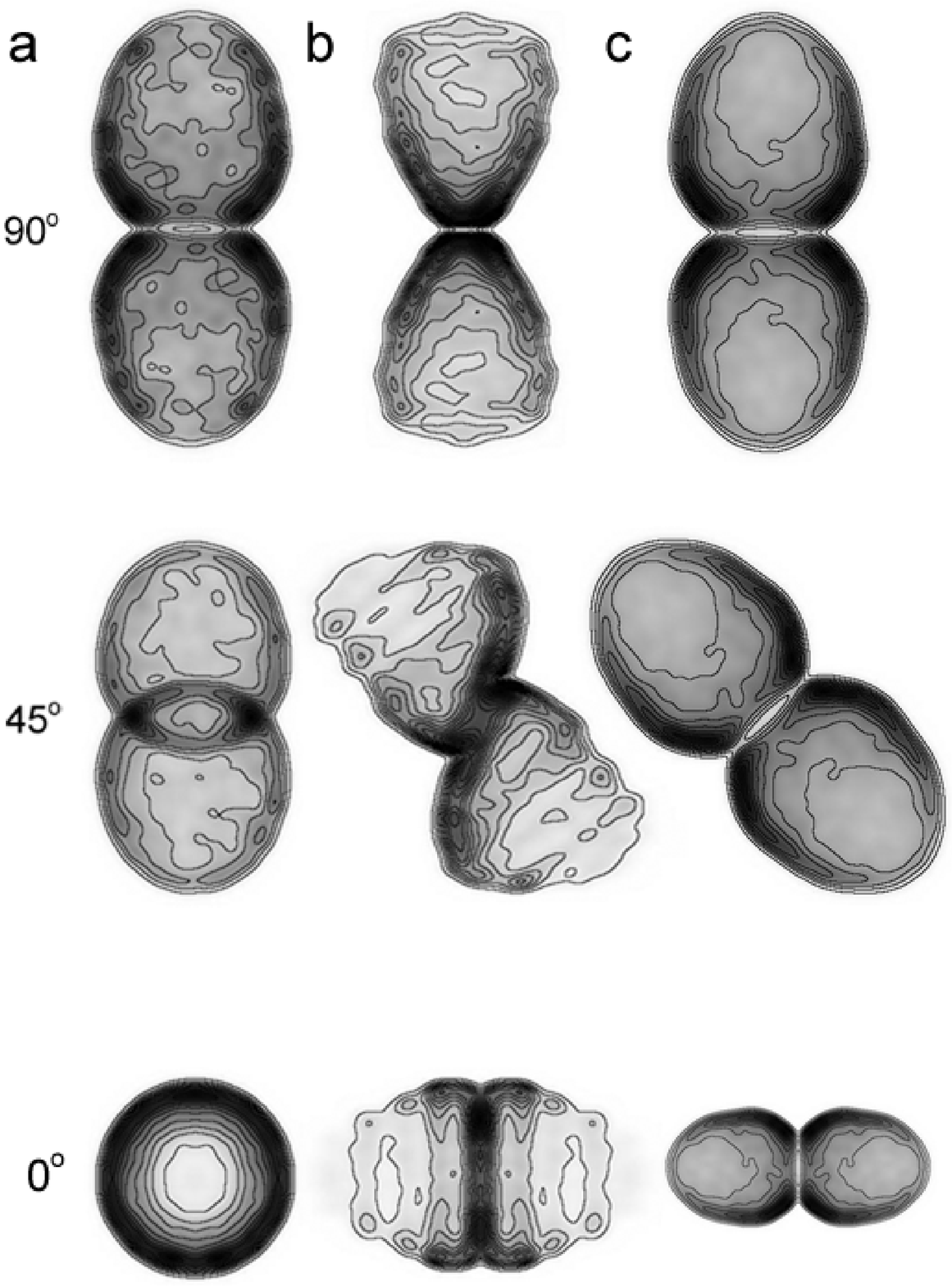}

  \caption{The rendered images of the hydrodynamical simulation B (column ''a''), slit-less spectra or homologous reconstruction (column ''b'') and the rendered
image of the reconstructed model with \shape (column ''c''). For further descriptions see Figure (\ref{fig:renders1}).}
  \label{fig:renders2}
  \vspace{0pt}

\end{figure}

\begin{figure}[!t]\centering

  \includegraphics[angle=00,width=0.95\columnwidth]{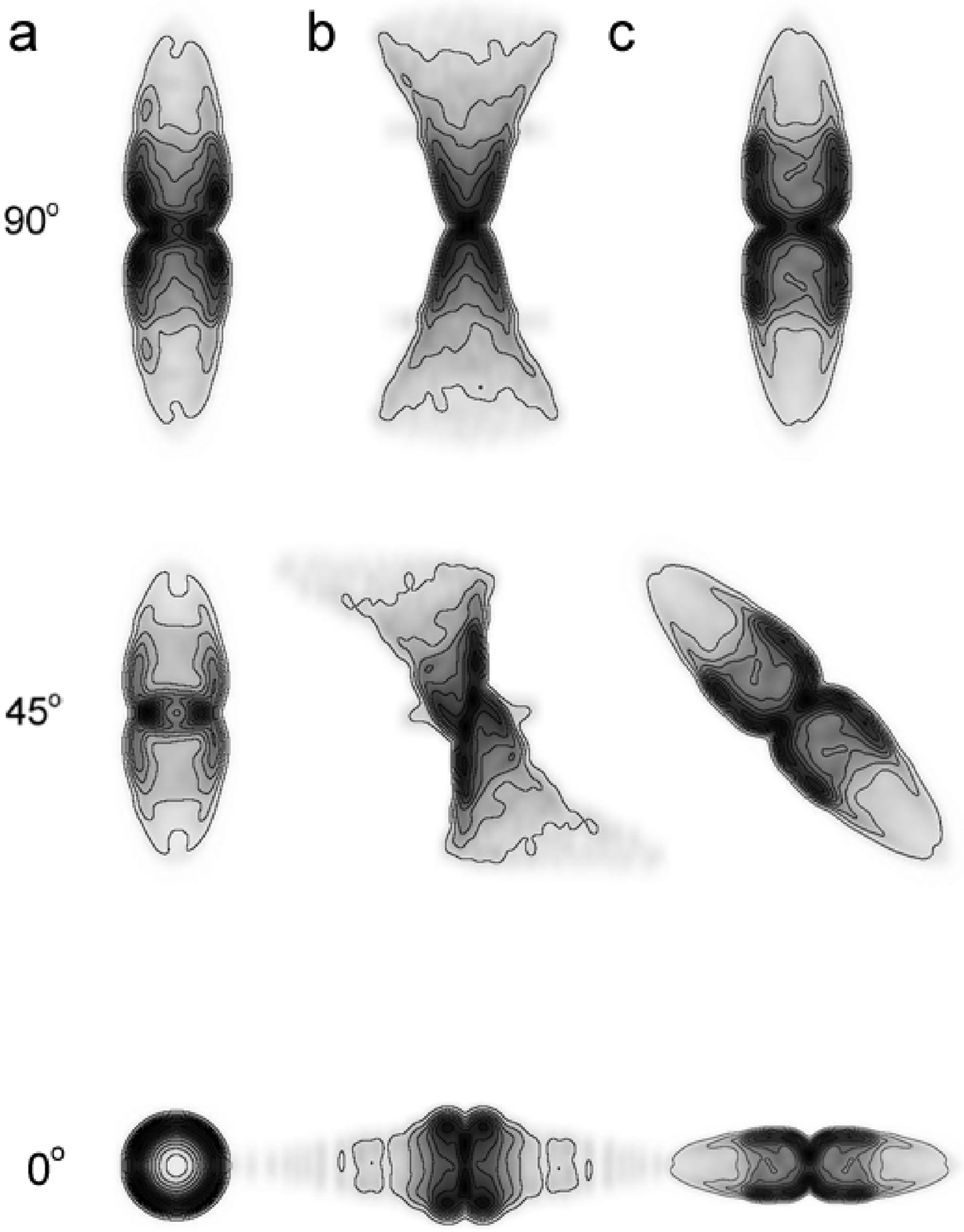}

  \caption{The rendered images of the hydrodynamical simulation Em (column ''a''), slit-less spectra or homologous reconstruction (column ''b'') and the rendered
image of the reconstructed model with \shape (column ''c''). For further descriptions see Figure (\ref{fig:renders1}).}
  \label{fig:renders3}
  \vspace{0pt}

\end{figure}

\begin{figure}[!t]\centering

  \includegraphics[angle=00,width=0.95\columnwidth]{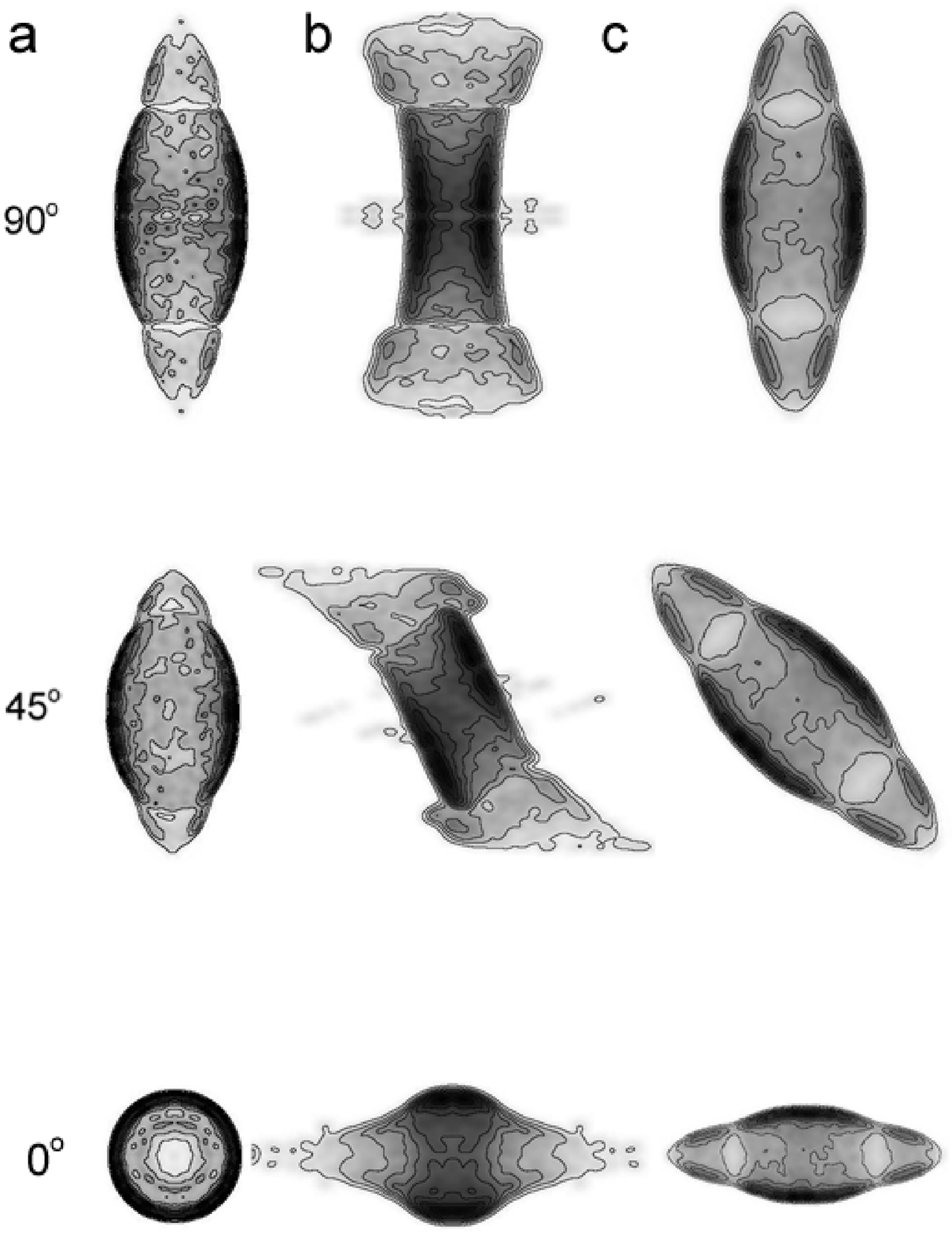}

  \caption{The rendered images of the hydrodynamical simulation Bm (column ''a''), slit-less spectra or homologous reconstruction (column ''b'') and the rendered
image of the reconstructed model with \shape (column ''c''). For further descriptions see Figure (\ref{fig:renders1}).}
  \label{fig:renders4}
  \vspace{0pt}

\end{figure}

Figures (\ref{fig:renders1}) to (\ref{fig:renders4})  show the rendered hydrodynamical simulations (columns ''a''),
their slit-less spectra or Hubble-like reconstructions (columns ''b'') and the reconstructed models with \shape (columns ''c'').
For the slit-less spectra, the simulated spectrograph slit was opened to include the complete object.
Note that a slit-less spectrum is equivalent to a 3D reconstruction that makes the assumption of a ''Hubble-type'' velocity distribution,
projected perpendicular to the line of sight.
A small gaussian blur has been applied to simulate observational seeing and reduce the stochastic particle-noise.
We discuss the results for each simulation separately.

\subsection{Non-MHD elliptical nebula}

Figures (\ref{fig:run1_2_r_vr}) and (\ref{fig:run1_2_dt}) show the statistics of the velocity component distribution
in the hydrodynamical simulation of run E which produced an ellipsoidal nebula (Figure \ref{fig:renders1}). The radial velocity component
increases approximately linearly with distance (Figure \ref{fig:run1_2_r_vr}). This might suggest that a homologous expansion is a good description of the velocity field. However,
the plot of the poloidal velocity component versus position angle
reveals that there are deviations of the velocity
vector from the radial direction (Figure \ref{fig:run1_2_dt}). The deviations increase from the
equator to poles until they reach approximately 7 km s$^{-1}$ at 20 $\deg$ from the pole. Close to the poles the poloidal velocity decreases again
to zero. Where the poloidal component is largest, noticeable deformations
will occur along the line of sight in a 3D reconstruction that assumes a homologous expansion law.

The image, the slit-less spectra and the reconstructed models are compared in Figure (\ref{fig:renders1}). The shape of
the slit-less spectrum  at $90^o$ (b)  is more rectangular than the direct image (a). However, the reconstructed model (c) fits
quite well. Although the deviations in (b) are small they are significant at intermediate to high latitudes.
For the $0^o$ and $90^o$ viewing angles, the qualitative difference is small and therefore might not influence significantly the physical interpretation of the
object. However, at intermediate viewing angles the P-V diagram is more distorted and becomes point-symmetric.
We consider it cautiously acceptable to directly reconstruct this type of objects from their spectral data assuming a homologous expansion law.

Both, the radial and the poloidal velocity show small-scale fluctuations which are associated with local instabilities. In a reconstruction, unresolved fluctuations will result in an increased thickness of the reconstructed surface structure.

\begin{figure}[!t]\centering

  \plotone{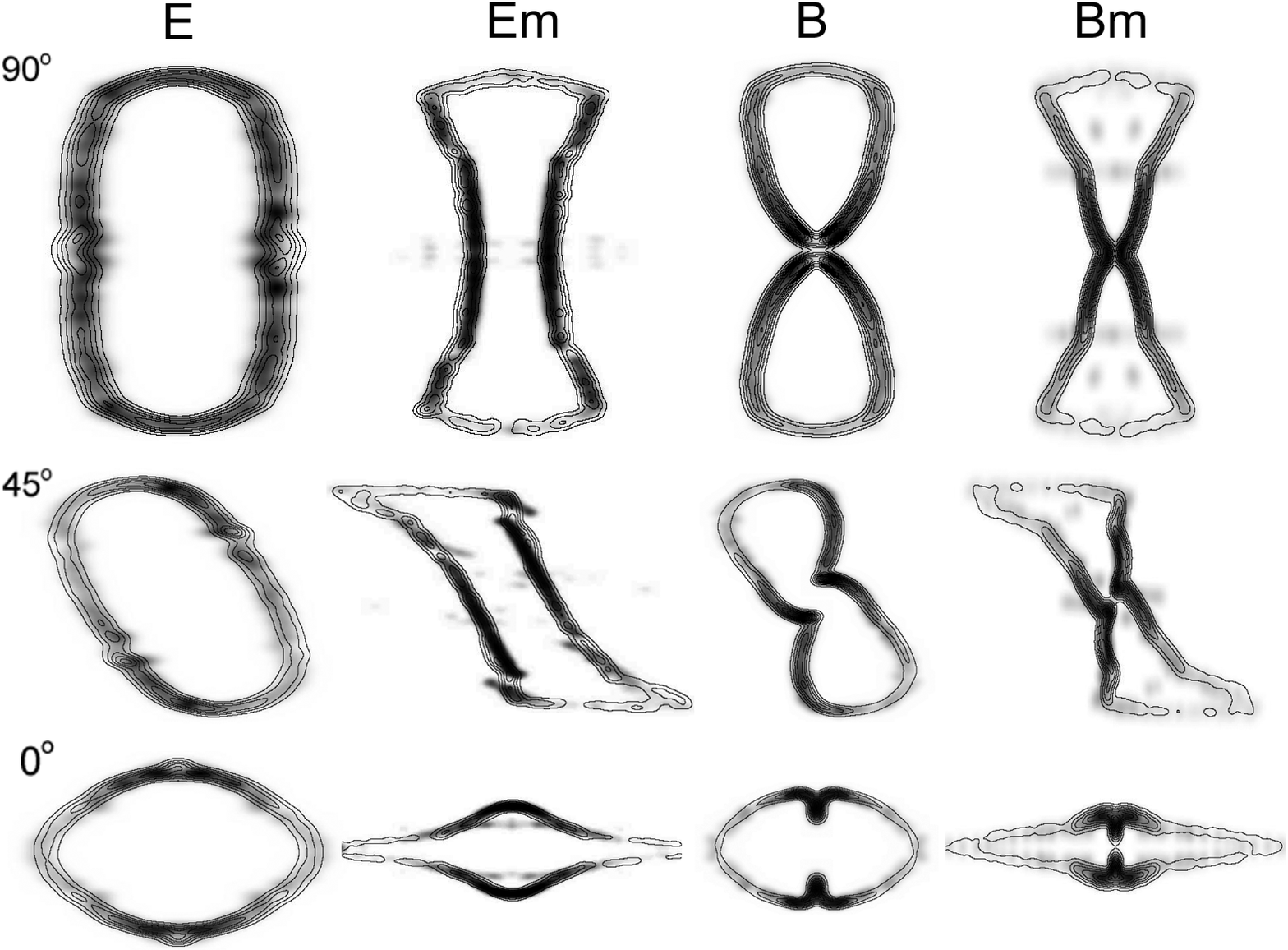}

  \caption{Central slit P-V diagrams from the simulations are represented in grey-scale. The superimposed contours
  are from the same simulations, but with a  simple piece-wise linear velocity distributions applied (dashed lines in Figures (\ref{fig:run1_2_r_vr}) to (\ref{fig:run3_4_dt}).}
  \label{fig:simple_model}
  \vspace{0pt}

\end{figure}

\subsection{Non-MHD bipolar lobes}

The radial velocity distribution for the bipolar lobes from the purely hydrodynamic model B is shown in Figure (\ref{fig:run1_2_r_vr}) and the distribution of the poloidal velocity is given in Figure (\ref{fig:run1_2_dt}). As in the ellipsoidal case of run E, we find a nearly linear increase of the radial velocity component with distance. Deviations from linearity are stronger far away and very close to the center. Also, the small-scale fluctuations increase at a certain distance. This distance marks the transition between two different dynamical regions in this bipolar nebula.

The general trend of the poloidal velocity is similar to the one in the ellipsoidal simulation, a linear increase with angle up to a maximum value and then a decrease towards the pole. The vanishing poloidal velocity at the pole and the equator are enforced in these simulations due to the boundary conditions of reflection at the equator and polar axis. However, it is expected that this behavior is very closely sustained in real objects with approximate axi-symmetry.

Column ''b'' in Figure (\ref{fig:renders2}) shows the slitless spectra of simulation B for 3 different viewing angles.
This corresponds to a 3D reconstruction that assumes homologous expansion as viewed along a direction perpendicular to the observer's line of sight. We note very significant differences to the observer's view in column ''a''. Especially, the ''Hubble-type reconstructions'' at $45^o$ and $0^o$ inclination angles show strong deviations. As in the case of run E, the intermediate angles introduce a point-symmetry which is now even stronger. Column ''c'' of Figure (\ref{fig:renders2}) shows the corresponding views of the axi-symmetric reconstructed
model using \shape which applies the simple piece-wise linear velocity law (see below).

In order to not only reproduce the images, but also the P-V diagrams with a
model that improves the homologous expansion law with only a few additional parameters,
we propose a simple piece-wise linear approximation to the velocity components as a
function of position.

For runs E and B, a two-parameter curve of two linear segments that start at zero at
the pole and equator and meet at a point ($\theta$, v$_{\theta}$) are sufficient.
The point ($\theta$, v$_{\theta}$) is a free parameter of the model. In addition,
a linear relation of the radial velocity with distance could be used. The result of
such a fit is shown in the P-V diagrams of Figure (\ref{fig:simple_model}, column E \& B).
The grey-scale P-V diagrams are from the hydrodynamical simulations with its original
velocity information. The contours are from the same structural data, but with the new
simple piece-wise linear velocity field. As an example, the corresponding velocity components
of run B are
plotted as dotted lines in Figures (\ref{fig:run1_2_r_vr}) \& (\ref{fig:run1_2_dt}).

We find that the agreement of this simple piece-wise linear velocity field with the
simulation is very satisfactory. At any viewing angle deviations in the P-V diagram
are only significant at the small scales of instabilities. Hence, for an elliptical
or bipolar hydrodynamical nebula a very significant improvement in its morpho-kinematical
modeling may be achieved by including only one and two linear segments in the distribution
of the radial and poloidal velocity components, respectively. This procedure retains the
axi-symmetric structure of the object and, simultaneously, provides a good fit to the
point-symmetric P-V diagrams.

\subsection{MHD collimated nebulae}
\label{sec:non-mhd}

We now perform a similar analysis to the MHD-simulations.
Figure (\ref{fig:run3_4_r_vr}) shows the distribution of the radial velocity
components of simulations Bm and Em as a function of distance from the center.
The poloidal velocity component is plotted in Figure (\ref{fig:run3_4_dt}). Both
simulations show similar velocity fields, but with different positions and amplitudes
of the dynamical regions.
The radial velocity component has a nearly linear increase up to about half of the
extent of the nebula. Then an abrupt change is noticeable with a small drop in the
velocity and further out a significant gradual increase in the slope of the velocity graph.

The poloidal velocity component is slowly growing from the equator towards the pole and
shows an accelerated increase from mid-latitudes on. It reaches
a maximum near the tip of the outflow, after which it declines rapidly
(as expected from the symmetry of the flow, see above). There are no strong
small-scale variations in either velocity component as opposed to those found in the
purely hydrodynamic simulations.

Figures (\ref{fig:renders3}) and (\ref{fig:renders4}) compare the projected images rendered from the MHD-simulations (column ``a'')
with the ``3D Hubble-type reconstructions'', i.e., the slit-less spectra (column ``b''),
and the reconstructed model using \shape (column ``c'').
The distortions observed in columns ''b'' are even stronger than in the non-MHD
simulations.
Instead of being closed bipolar lobes, they appear to be open. In all cases of
intermediate inclination angles, the axi-symmetry of the simulation is lost and replaced
by point-symmetry. A physical interpretation of such artificial symmetry features, in
3D reconstructions that assume homologous expansion, may lead to seriously erroneous
results.

In order to capture the main features of the velocity field in this type of objects in
morpho-kinematical modeling, we propose to use at least two linear segments for the
radial component and three segments for the poloidal velocity component. Such
approximations for model Bm have been plotted in Figures (\ref{fig:run3_4_r_vr}) and
(\ref{fig:run3_4_dt}) as dotted lines. See Figure (\ref{fig:simple_model}) to
compare the P-V diagrams from the simplified \shape model with those of the simulation.
The deviations from the simulation are very small, hence, also for the MHD cases a
considerable improvement of the velocity field can be obtained by the simple piece-wise
linear velocity field proposed here.

\begin{figure}[!t]\centering

  \includegraphics[angle=00,width=0.45\columnwidth]{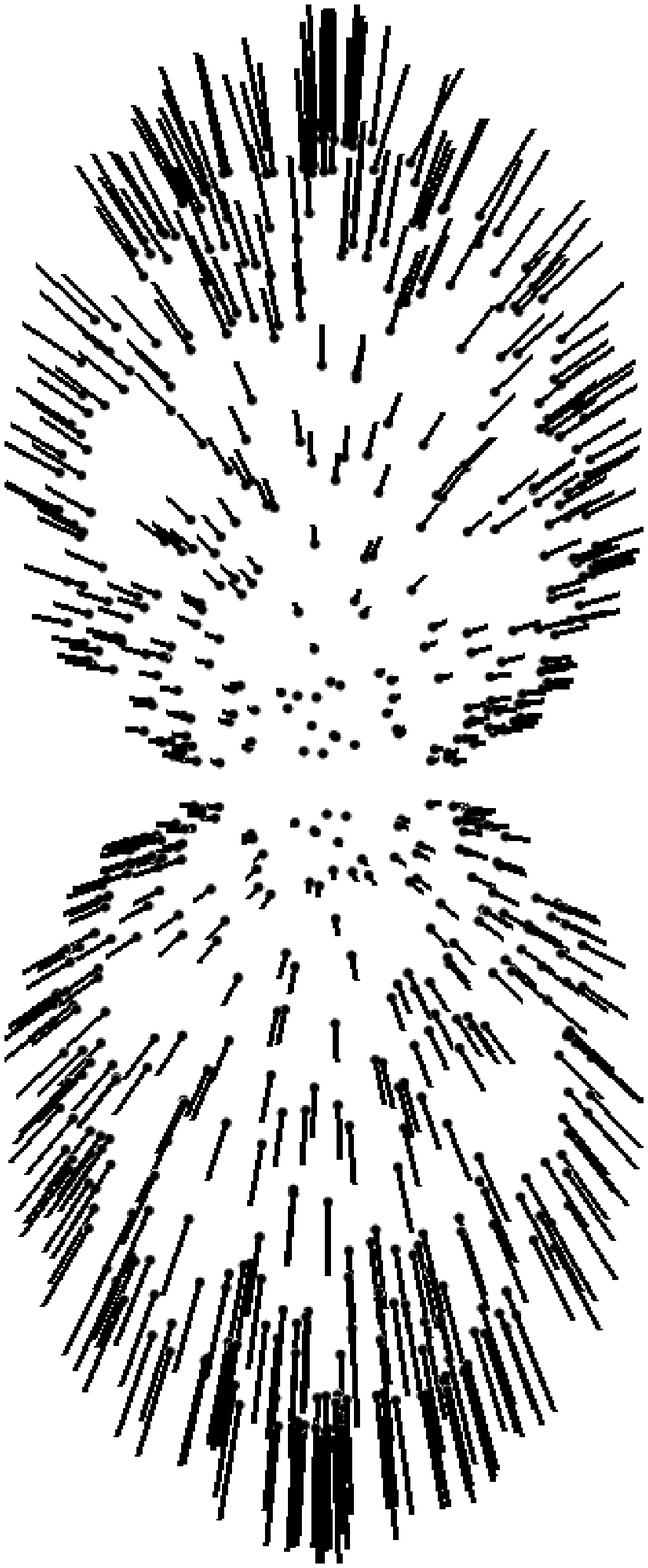}
  \includegraphics[angle=00,width=0.45\columnwidth]{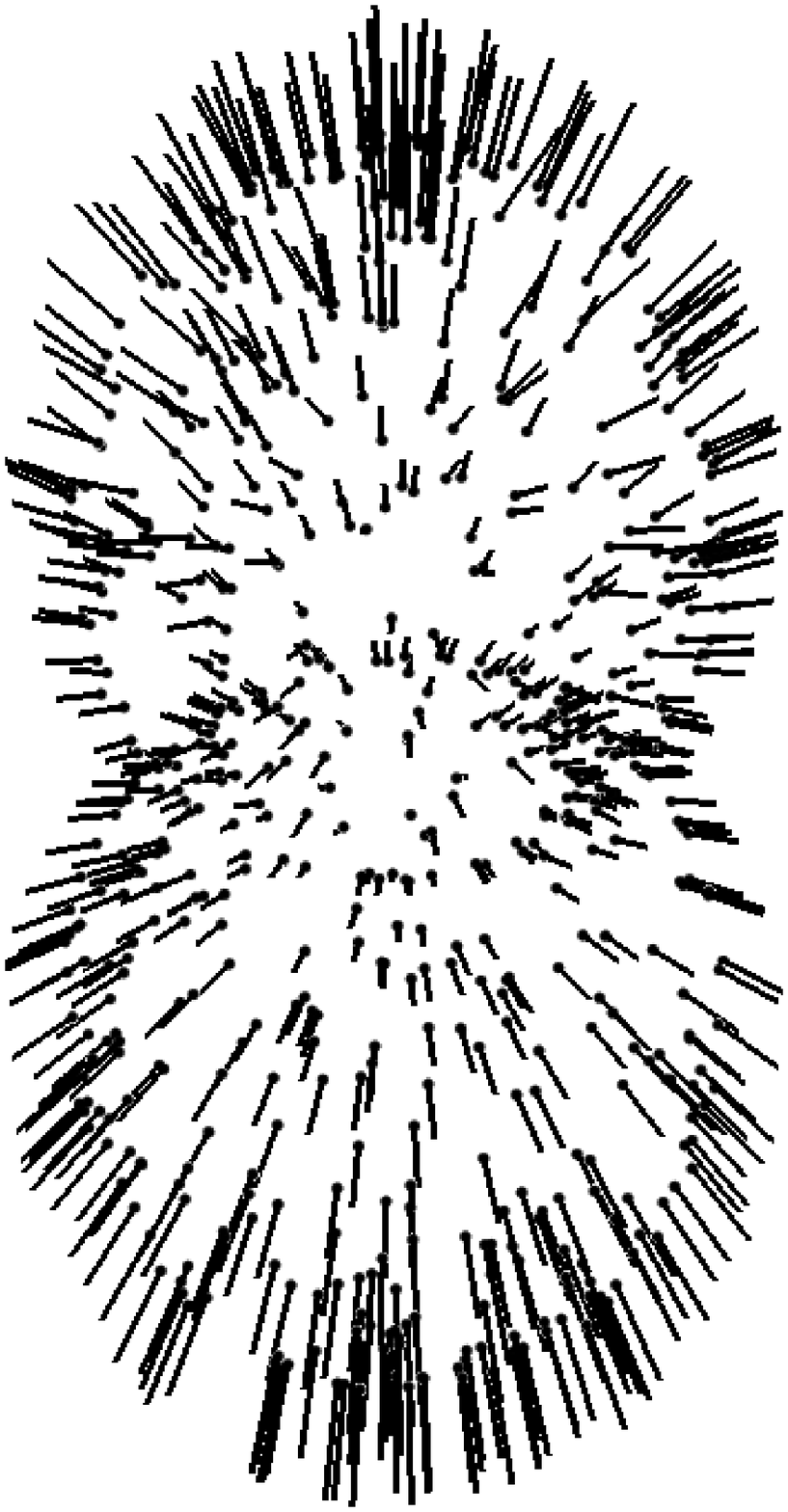}
  \caption{Proper motion vectors are shown for model B at 90 and 45 degrees viewing angle on the left and right, respectively. The top half has the velocity field as obtained from the simulation, while the bottom has a homologous expansion law.}
  \label{fig:proper_motion_1}
  \vspace{0pt}

\end{figure}

\begin{figure}[!t]\centering

  \includegraphics[angle=00,width=0.45\columnwidth]{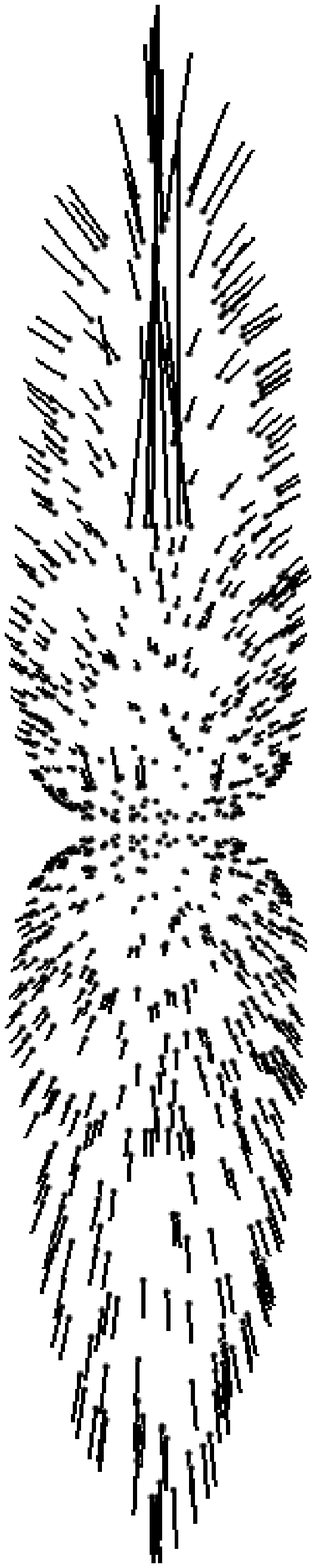}
  \includegraphics[angle=00,width=0.45\columnwidth]{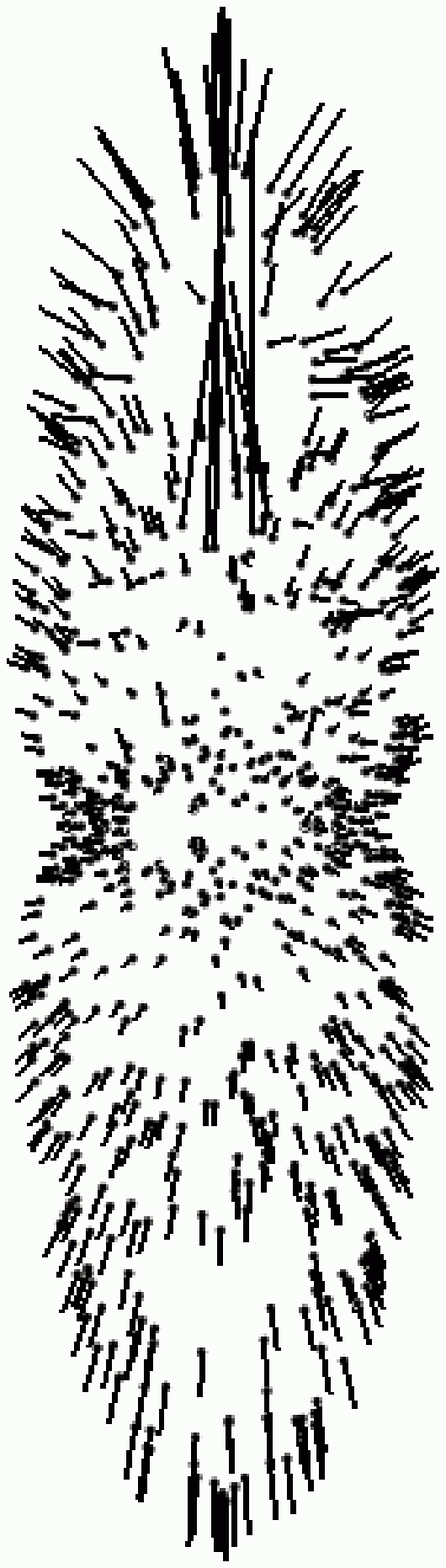}
  \caption{Proper motion vectors are shown for model Bm at 90 and 45 degrees viewing angle on the left and right, respectively. The top half has the velocity field as obtained from the simulation, while the bottom has a homologous expansion law. The top half includes part of the central collimated outflow with very large velocity vectors.}
  \label{fig:proper_motion_3}
  \vspace{0pt}

\end{figure}

\section{Proper motion patterns}
\label{sec:proper_motion}

Additional kinematic information from an expanding object can be obtained from internal
proper motion measurements. Such measurements have been obtained for a few nearby
planetary nebulae (e.g. NGC6302: Meaburn et al. 2008; NGC7009: Rodr\'{\i}guez \& G\'omez
2007). Measurements of Doppler-shift and proper motion of a large number of features in a
nebula (e.g. Li, Harrington \& Borkowski 2002) might reveal whether an object is
hydrodynamically active or expanding homologously due to ballistic motion.
In Figures (\ref{fig:proper_motion_1}) and (\ref{fig:proper_motion_3}) we compare the
proper motion vectors of our simulations B and Bm with the corresponding homologous
motion patterns for two different viewing angles. The top halves of the images represent
the velocity pattern from the hydrodynamical simulations, while in the bottom halves we
replaced the original velocity field by a homologous velocity field. Clearly, such
measurements might reveal a lot of information about the dynamical state of a planetary
nebula. When the expansion is homologous, the proper motion pattern reflects this in
purely radial vectors with magnitudes that increase linearly with distance.
This pattern is basically independent of the structure and orientation of the object.

However, in a dynamically active nebula or in a nebula with several kinematic sub-systems,
the motion pattern will depend on the orientation and structure. Some of the proper
motion vectors in Figures (\ref{fig:proper_motion_1}) and (\ref{fig:proper_motion_3})
almost reverse direction in the inclined objects (right side). Such a behavior is found
in BD+30$^{o}$3639. As seen from the proper motion pattern in Li, Harrington \&
Borkowski (2002), BD+30$^{o}$3639 is likely to still be a dynamically active nebula.
The spectral and proper motion data from Meaburn et al. (2008) indicate that
NGC~6302 is expanding radially and, hence, presumably ballistically.

In general, because of the evolution of the fast stellar wind, it is expected that
young planetary nebulae will be more likely to be active than older nebulae. It is
therefore surprising that very young non-spherical nebulae, like, for instance,
Minkowski's footprint (M1-92, Alcolea, Neri \& Bujarrabal 2007) show Doppler-shift
patterns that are consistent with axi-symmetric homologous expansion.
It is important to systematically survey the kinematic properties of planetary nebulae
for evidence of deviations from homologous expansion. This will reveal information
about the evolution of the fast stellar wind and the evolution of central stars
in general. Hopefully, the analysis of large kinematical catalogs like those of Hajian et al. (2007) and L\'opez et al. (2006) might be helpful in studying the transition phase between hydrodynamically active and ballistic expansion.

\section{Conclusions}

Based on axi-symmetric hydrodynamical simulations and 3D reconstructions with
the software package {\it Shape}, we have investigated the kinematic
signatures of deviations from ``Hubble-type'' outflows in some typical shapes of planetary nebulae.
The deviations are systematic and their description as
given in this paper may
help to improve morpho-kinematic and dynamic models of planetary nebulae.
Excluding the small-scale variations from local instabilities,
we find that the radial velocity as a function of distance can be well described by a
linear increase for the cases of purely hydrodynamical ellipsoidal and bipolar-lobed
nebulae. The poloidal component is, however, not negligable. Its parameterized
approximation requires two linear segments, one increasing and the other decreasing
with zero velocity at the pole and equator.
We also find that the magneto-hydrodynamically generated cases require two linear
segments for the radial velocity component and three segments for the poloidal component.
These simple extensions to a homologous expansion law capture the global velocity
structure of hydrodynamically active axi-symmetric nebulae except for small-scale
variations.

We find that, in most situations considered in our simulations,
the deviations from a homologous flow, both in magnitude and
direction of the velocity vector, are significant and
observable. The only case where only small deviations were
found was the case of an ellipsoidal nebula.

If significant deviations from a homologous expansion model can
be inferred, then it is very likely that it is still
dynamically active in the sense that the fast wind is still significant or
the inner region of the nebula is still over-pressurized  with respect to
the exterior. In a future paper we will explore more precisely how
the transition from the hydrodynamically to the ballistic regimes occurs.

It is the relative
size of the poloidal velocity component that strongly influences the
shape of the position velocity diagrams that are obtained,
not so much the variation of the radial component with distance from the center.
The deviations increase with the degree of collimation of the nebula.
They are stronger at intermediate latitudes.
There is a slow increase in deviations as we move from the equator to
the poles, usually with a sudden increase at mid-latitudes.

A good description of the velocity field in the MHD-simulations requires one more
linear segment than that of the non-MHD simulations. This would imply that
this difference might provide observational evidence for the presence or absence
of dynamical active magnetic fields in the formation process of planetary nebulae.
However, this aspect might require more detailed and specific studies.

Summing up, detailed morpho-kinematic observations and modeling of planetary
nebulae can reveal whether a nebula is still in a hydrodynamically active stage
or whether it has reached ballistic expansion.

\acknowledgments

We thank Michael L. Norman and the Laboratory for Computational
Astrophysics for the use of ZEUS-3D. We also thank J.A. L\'opez for his very
useful discussions on the observed properties of planetary nebulae and their
interpretation.  We acknowledge financial support from CONACYT grant 49447 and
UNAM DGAPA-PAPIIT  IN108506-2. We thank the anonymous referee for useful suggestions.

\end{document}